\documentclass[twocolumn,superscriptaddress,showpacs,aps]{revtex4}

\usepackage{mathrsfs}
\usepackage{hyperref}

\usepackage[dvips]{graphics,epsfig}
\usepackage{amstext}
\usepackage{amssymb}
\usepackage{amsmath}
\usepackage{graphicx,color}
\usepackage{bm}

\newcommand{\s}{\mbox{\tiny S}}
\newcommand{\B}{\mbox{\tiny B}}
\newcommand{\SB}{\mbox{\tiny SB}}
\newcommand{\T}{\mbox{\tiny T}}
\newcommand{\M}{\mbox{\tiny M}}

\newcommand{\wti}{\widetilde}
\newcommand{\nl}{\nonumber \\}

\newcommand{\App}[1]{Appendix\,\ref{#1}}
\newcommand{\be}{\begin{equation}}
\newcommand{\ee}{\end{equation}}
\newcommand{\bea}{\begin{eqnarray}}
\newcommand{\eea}{\end{eqnarray}}
\newcommand{\bsube}{\begin{subequations}}
\newcommand{\esube}{\end{subequations}}
\newcommand{\Eq}[1]{Eq.\,(\ref{#1})}
\newcommand{\Eqs}[1]{Eqs.\,(\ref{#1})}
\newcommand{\Fig}[1]{Fig.\,\ref{#1}}

\newcommand{\dg}{\dagger}
\newcommand{\la}{\langle}
\newcommand{\ra}{\rangle}

\newcommand{\mb}{\mbox}

\begin{document}

\title{Master equation approach for transport through Majorana zero modes }

\author{Jinshuang Jin} \email{jsjin@hznu.edu.cn}
\affiliation{School of Physics, Hangzhou Normal University,
Hangzhou, Zhejiang 311121, China}

\author{Xin-Qi Li} \email{xinqi.li@tju.edu.cn}
\affiliation{Center for Joint Quantum Studies and Department of Physics,
School of Science, Tianjin University, Tianjin 300072, China}

\date{\today}

\begin{abstract}
Based on an exact formulation,
we present a master equation approach to
transport through Majorana zero modes (MZMs).
Within the master equation treatment, the occupation dynamics
of the regular fermion associated with the MZMs
holds a quite different picture
from the Bogoliubov–de Gennes (BdG)
$S$-matrix scattering process, in which the ``positive"
and ``negative" energy states are employed,
while the master equation treatment does not involve them at all.
Via careful analysis for the structure of the rates
and the rate processes governed by the master equation,
we reveal the intrinsic connection between both approaches.
This connection enables us to better understand the confusing issue
of teleportation when the Majorana coupling vanishes.
We illustrate the behaviors of transient rates, occupation dynamics and currents.
Through the bias voltage dependence,
we also show the Markovian condition for the rates,
which can extremely simplify the applications in practice.
As future perspective, the master equation approach developed in this work
can be applied to study important time-dependent phenomena
such as photon-assisted tunneling through the MZMs
and modulation effect of the Majorana coupling energy.
\end{abstract}

\maketitle 

\section{Introduction}
\label{thintro}

Majorana zero modes (MZMs)
obey non-Abelian statistics and have sound potential for
topological quantum computations \cite{Kit01131,Kit032,Nay081083,Lei11210502}.
The practical realization and identification of MZMs have thus received great interest
in both theoretical \cite{Lut10077001,Ore10177002,
Fle10180516,Sar16035143,Vui19061,Wu12085415,Moo18165302,Law09237001,Nil08120403,
Dan20036801,Men20036802}
and experimental studies
\cite{Mou121003,Den126414,Das12887, 
Fin13126406,Alb161038}.
For identification, existing proposals for transport measurements
were largely based on tunneling conductance spectroscopy,
employing either a single-lead (two-terminal) setup
\cite{Fle10180516,Sar16035143,Vui19061,Wu12085415,Moo18165302,Law09237001}
or a more powerful two-lead (three-terminal) setup
\cite{Nil08120403,Wu12085415,Moo18165302,Law09237001,
Dan20036801,Men20036802}.
However, it has been argued that both the zero-bias conductance peak
and its quantized value $2e^2/h$
are not the ultimate decisive evidences of identifying the existence of MZMs.
A key task remaining in this field is to distinguish the MZMs
from other possible bound states, e.g., the Andreev bound states,
which can also result in similar zero-bias conductance peak.

In addition to the steady-state transport study for such as the zero-bias conductance peak,
future study may turn more attention to dynamical aspects associated with the MZMs,
which may include problems such as motionally moving the MZMs,
modulating the coupling between the MZMs,
and photon-assisted tunneling through MZMs, etc.
In particular, one may consider to insert all these studies
into a transport context,
which is anticipated to make the unique dynamics of the MZMs
measurable through time-dependent transport currents.
For isolated Majorana systems,
the theoretical tool of quantum dynamics simulation
can be the time-dependent Schr\"odinger equation.
However, for the open system of MZMs coupled to outside environments
(e.g. fluctuating background charges and/or transport leads),
it is required, usually, to develop a master equation approach
such as done in the recent works \cite{Lai18054508,Hua20165116},
where the dissipative quantum dynamics caused by local noises,
 transient evolution of the zero-bias conductance peak,
and finite bandwidth effect of the leads are numerically investigated.

Indeed, master equation approach is an alternative useful tool
for studying transport through mesoscopic systems
\cite{Bru941076,Bru974730,Gur9615932,Leh02228305,Li05066803,Li05205304,Luo07085325,Li11115319,
Jin08234703,Jin10083013},
aside from the well-known Landauer-B\"uttiker scattering matrix theory \cite{Dat95}
and the nonequilibrium Green's function formalism \cite{Hau08}.
One of the distinct advantages of the master equation approach
is its convenience for studying
the various statistical properties of transport currents.
In transport through the MZMs,
the charge transmission channels contain normal and Andreev processes.
In steady state, the Bogoliubov–de Gennes (BdG)
$S$-matrix scattering approach
can describe the various channels very clearly.
However, in the master equation approach,
which describes the occupation dynamics
and allows for current calculation,
it is unclear how to infer these transport channels.
In this work, we carry out explicit results
for the master equation and currents of transport through a pair of MZMs.
In particular, via careful analysis for the structure of the rates
involved in the master equation,
we will reveal the intrinsic connection between
the master equation and the BdG $S$-matrix scattering approach.
To the best of our knowledge, this important connection is absent in literature.
This connection can also help us to better understand
the confusing issue of teleportation when the Majorana coupling vanishes.

The work is organized as follows.
In Sec.\ II we present the main results derived in Appendix A,
including the master equation and time-dependent currents.
In Sec.\ III we carry out numerical results for the transient rates and currents,
illustrate the bias condition for Markovian approximation,
and analyze the steady-state results in detail
by a connection with the BdG $S$-matrix scattering approach.
Finally, summary and discussion are presented in Sec.\ IV.

\begin{figure}
\includegraphics[width=0.7\columnwidth]{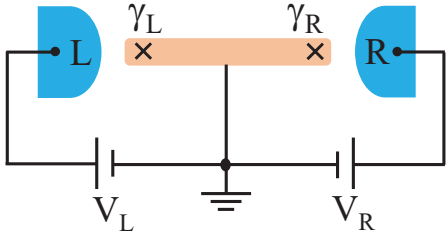} 
\caption{(Color online)
Sketch of the two-lead (three-terminal) setup
for transport through a pair of MZMs
($\gamma_{\rm L}$ and $\gamma_{\rm R}$, with coupling energy $\varepsilon_{\M}$).
The two leads are biased by voltages $V_{\rm L}$ and $V_{\rm R}$
with respect to the Fermi level of the superconductor,
while the superconductor is grounded through the third terminal.       }
\label{fig1}
\end{figure}

\section{Formulation}

For the transport setup of a pair of MZMs coupled to transport leads,
the total Hamiltonian consists of three parts,
\be
H_{\T}=H_{\s}+H_{\B}+H_{\SB}.
\ee
The cental TS quantum wire can be described by the low-energy effective
Hamiltonian of a pair of MZMs,
$H_{\s}=\frac{i}{2} \varepsilon_{\M} \gamma_{\rm L}\gamma_{\rm R}=
\varepsilon_{\M}(f^\dg f-\frac{1}{2})$,
where $\varepsilon_{\M}$ is the coupling energy
between $\gamma_{\rm L}$ and $\gamma_{\rm R}$.
The two Majorana fermions are related to a regular fermion
through the transformation
of $\gamma_{\rm L}= (  f+  f^\dg)$ and $\gamma_{\rm R}= -i(  f-  f^\dg)$, with
$\{\gamma_{\alpha},\gamma_{\beta}\}=2\delta_{\alpha\beta}$, $\gamma^\dg_{\alpha}=\gamma_{\alpha}$,
and $\gamma^2_{\alpha}=1$.
The transport leads can be modeled by the noninteracting electron Hamiltonian as
$H_{\B}=\sum_{\alpha k}\varepsilon_{\alpha k} c^\dg_{\alpha k}c_{\alpha k}$,
with 
$c^\dg_{\alpha k} (c_{\alpha k})$ the creation (annihilation) operators of the electron
in the $\alpha$-lead ($\alpha={\rm L/R}$, the left/right lead).
The tunnel-coupling between the MZMs and the two leads is described
by \cite{Zaz11165440,Fle10180516,Cao12115311},
 \begin{align}\label{HSB0}
  H_{\SB}=\sum_{  k}  \big(t_{L k}c^\dg_{L k}\gamma_{\rm L}+
   it _{\rm R k}c^\dg_{R k}\gamma_{\rm R}+{\rm H.c.}\big).
 \end{align}
Throughout this work, we set $\hbar=1$, unless otherwise stated.

\subsection{ Exact Formulation of master equation}

Let us rewrite the tunnel-coupling Hamiltonian \Eq{HSB0} as
\be\label{HSB2}
 H_{\SB}=\sum_{\alpha}\big(F^\dg_{\alpha} Q_\alpha+{\rm H.c.}\big) \,.
 \ee
Here we denoted
$Q_{\rm L}\equiv \gamma_{\rm L}=(f+f^\dg)=Q^\dg_{\rm L}$,
$Q_{\rm R}\equiv i\gamma_{\rm R}=f-f^\dg=-Q^\dg_{\rm R}$,
and $F^\dg_{\alpha}\equiv\sum_kt_{ \alpha k}c^\dg_{ \alpha k} $.
For the convenience of latter presentation,
let us also introduce the correlation functions
$g^+_{\alpha}(t-\tau) = \la F^{\dagger}_{\alpha}(\tau)F_{\alpha}(t)\ra_{\B} $
and $g^-_{\alpha}(t-\tau) = \la F_{\alpha}(t)F^{\dagger}_{\alpha}(\tau)\ra_{\B}$,
where $\la\cdots \ra_B$ means the thermal average over the reservoir states of the leads.
Straightforwardly, one can more explicitly obtain
$ g^{\pm}_{\alpha}(t-\tau)
 =\sum_k e^{-i\varepsilon_{\alpha k}(t-\tau)}\,|t_{\alpha k}|^2
 n^{\pm}_\alpha(\varepsilon_{\alpha k})$,
where $n^{\pm}_\alpha$ are the Fermi occupied and empty functions.
The sum of these two functions reads as
$g_{\alpha}(t-\tau)  =\sum_k e^{-i\varepsilon_{\alpha k} (t-\tau)}\,|t_{\alpha k}|^2$.
Through the Fourier transformation
\bsube\label{corr-bath}
 \begin{align}
    g_{\alpha}(t-\tau)
 &=
 \int^\infty_{-\infty} \frac{d\omega}{2\pi}\,e^{-i\omega (t-\tau)}\,  \Gamma_\alpha(\omega) \,,
 \label{corrg-bath}
 \\
    g^{\pm}_{\alpha}(t-\tau)
 &=
 \int^\infty_{-\infty}\frac{d\omega}{2\pi}\,e^{-i\omega (t-\tau)}\,  \Gamma^{\pm}_\alpha(\omega) \,,
 \end{align}
 \esube
we also introduce the corresponding rates
$\Gamma_{\alpha}(\omega)=  2\pi \sum_k
|t_{\alpha k}|^2\delta(\omega-\varepsilon_{\alpha k})$
and $\Gamma^{\pm}_\alpha(\omega) = \Gamma_\alpha(\omega)n^{\pm}_\alpha(\omega)$.

The basic idea of master equation approach to quantum transport
is regarding the transport leads as a generalized environment
and constructing an equation-of-motion for
the reduced state of the central device system,
$\rho(t)\equiv {\rm tr}_{\B}\rho_{\T}(t)$,
where the trace is over the states of the transport leads
and $\rho_{\T}(t)$ is the total density operator of the device-plus-leads.
The starting equation is the Schr\"odinger equation for the total wavefunction,
or, equivalently, the Liouvillian equation for $\rho_{\T}(t)$,
$\dot{\rho}_{\T}(t)=-i[H_{\T},\rho_{\T}(t)]$.
Performing the trace ${\rm tr}_{\B}(\cdots)$ on two sides of the Liouvillian equation,
one obtains
\begin{align}\label{QME0}
\dot{\rho}(t)&\!\!=\!\!-i[H_{\s},\rho(t)]\!-\!i\!\sum_{\alpha}[Q_\alpha,\rho^{+}_{\alpha}\!(t)]
\!-\!i\!\sum_{\alpha}[Q^\dg_\alpha,\rho^{-}_{\alpha}\!(t)],
\end{align}
where the two auxiliary density operators (ADOs) are introduced as
\begin{align}\label{rhoaux}
 \rho^+_{\alpha }(t)\!\equiv\!\mb{tr}_{\B}\{\rho_{\T}(t)F^\dg_\alpha \}~\text{and}~
 \rho^-_{\alpha }(t)\!\equiv\!\mb{tr}_{\B}\{F_\alpha \rho_{\T}(t)\} \,.
  \end{align}
They satisfy $\rho^+_{\alpha }(t)=[\rho^-_{\alpha}(t)]^\dg$.
By means of a path-integral formulation
in the fermionic coherent state representation \cite{Jin10083013},
it is possible to express the ADOs $\rho^{\pm}_{\alpha }(t)$
in terms of the reduced density operator $\rho(t)$.
For the MZMs system under study,
the detailed derivation is presented in Appendix A   
and the final result reads as
\begin{align}\label{EME-wbl}
\dot{\rho}(t)&=
-i[  H_{\s}(t),\rho(t)]+\Gamma_{1}(t)
{\cal D}[  f]\rho(t)
+\Gamma_{2}(t)
{\cal D}[  f^\dg]\rho(t)
\nl&\quad
 +\Upsilon(t)\big(  f\rho   f+ f^\dg\rho f^\dg\big)  \,,
\end{align}
where the Lindblad superoperator is defined by
${\cal D}[A]\rho=A\rho A^\dg-\frac{1}{2}\{A^\dg A,\rho\}$
and the time-dependent rates are given by
\bsube\label{Gama12t}
\begin{align}
\Gamma_{1}(t)&\equiv\sum_{\alpha}\big[\Gamma^-_{\alpha 11}(t)+\Gamma^+_{\alpha 22}(t)\big]
\nl&\quad
-\sum_{\alpha}(-)^{\alpha}\big[ \Gamma^-_{\alpha 12}(t)+ \Gamma^+_{\alpha 21}(t)\big] \,,
\label{Gam1t}
\\
\Gamma_{2}(t)&\equiv
\sum_{\alpha}\big[\Gamma^+_{\alpha 11}(t)+\Gamma^-_{\alpha 22}(t)\big]
\nl&\quad
-\sum_{\alpha}(-)^{\alpha}\big[ \Gamma^+_{\alpha 12}(t)+ \Gamma^-_{\alpha 21}(t)\big] \,,
\label{Gam2t}
\\
\Upsilon(t)&= \Gamma_{\rm L}(t)-\Gamma_{\rm R}(t) \,.
\end{align}
\esube
Assuming a wide-band limit approximation for the transport leads,
i.e., $\Gamma_{\alpha}(\omega) \rightarrow\Gamma_{\alpha}$ in \Eq{corr-bath},
the various individual rates can be more simply expressed as
\bsube\label{rates-wbl}
\begin{align}
\Gamma^{+}_{\alpha 11} (t)&
\!=\!2\,{\rm Re}\!\int_{t_0}^{t}\!d\tau\,
     g^+_{\alpha}(t-\tau)u^\dg_{11}(t,\tau),
     \label{Gamap11}
\\
\Gamma^{-}_{\alpha 22} (t)&
  \!=\!2\,{\rm Re}\!\int_{t_0}^{t}\!d\tau\,g^-_{\alpha}(t-\tau)u_{11}(t,\tau),
   \label{Gaman22}
\\
 \Gamma^{+}_{\alpha 12} (t)&
 \!=\!2{\rm Re}\int_{t_0}^{t}\!d\tau\,g^+_{\alpha}(t-\tau)u^\dg_{12}(t,\tau),
\label{Gamap12}
\\
 \Gamma^{-}_{\alpha 21} (t)&
 \!=\!2{\rm Re}\int_{t_0}^{t}\!d\tau\,
      g^-_{\alpha}(t-\tau)u_{12}(t,\tau),
 \label{Gaman21}
\\
\Gamma_{\alpha}(t)&\!=\! 2\,{\rm Re}\!\int_{t_0}^{t}\!d\tau\,g_{\alpha}(t-\tau)  \,,
\label{Gammalead}
\end{align}
while the other rates can be obtained through the following relations
\begin{align}
\Gamma^{-}_{\alpha 11}(t)
  & =\Gamma_{\alpha}(t)
- \Gamma^{+}_{\alpha 11}(t)  \,,
\label{Gaman11}
\\
\Gamma^{+}_{\alpha 22}(t)
  & =\Gamma_{\alpha}(t)
- \Gamma^{-}_{\alpha 22} (t) \,,
\label{Gamap22}
\\
\Gamma^{-}_{\alpha ij} (t)&=-\Gamma^{+}_{\alpha ij} (t), (i\neq j) \,.
\label{Gamanp12}
\end{align}
\esube
In the above results, the {\it reduced} propagators (RPs)
for the central MZMs system $u_{ij}(t,\tau)$,
in terms of a matrix form $\bm u(t,\tau)$,
satisfy the following equation-of-motion
\begin{align}
\label{ueq1}
\frac{d}{dt}{\bm u}(t,t_0)+\!\left(
                                  \begin{array}{cc}
                                    i\varepsilon_{\M} +\Gamma
                                    & \Gamma_{\rm d}\\
                                    \Gamma_{\rm d}
                                    & -i\varepsilon_{\M} +\Gamma \\
                                  \end{array}
                                \right){\bm u}(t,t_0)\!=\!0 \,.
\end{align}
Here we also considered the wide-band limit and introduced that
$\Gamma=\Gamma_{\rm L}+\Gamma_{\rm R}$ and $\Gamma_{\rm d}=\Gamma_{\rm L}-\Gamma_{\rm R}$.
Actually, ${\bm u}(t,t_0)$ is nothing but the superconductor Green's function (GF) matrix,
specified here for the MZMs coupled the transport leads.
Therefore, for the normal GFs (the diagonal elements), we have
$u_{11}(t_0,t_0)=u_{22}(t_0,t_0)=1$;
and for the anomalous GFs (the off-diagonal elements), we have
$u_{12}(t_0,t_0)=u_{21}(t_0,t_0)=0$.
Moreover, we have ${ u}^\dg_{11}(t,t_0)={ u}_{22}(t,t_0)$
and ${ u}^\dg_{12}(t,t_0)={ u}_{21}(t,t_0)$.

Owing to the symmetry properties of the ${\bm u}(t,t_0)$ matrix,
from \Eq{Gamap12} and (\ref{Gaman21}),
one can prove that $\Gamma^{+}_{\alpha 12} (t)=-\Gamma^{-}_{\alpha 21} (t)$.
Together with \Eq{Gamanp12},
we then have $\Gamma^{\pm}_{\alpha 12} (t)=\Gamma^{\pm}_{\alpha 21} (t)$.
Remarkably, for the master equation, the total rates
$\Gamma_1(t)$ and $\Gamma_2(t)$ of \Eq{Gama12t}
are determined only by the sum of the diagonal rates $\Gamma^{\pm}_{\alpha 11/22} (t)$.
The off-diagonal rates $\Gamma^{\pm}_{\alpha 12/21} (t)$
have no effect on the state evolution.
However, as we will see later,
the off-diagonal rates have important contribution to the transport currents.

So far, we have presented the main results of the master equation,
i.e., \Eq{EME-wbl} and the various rates associated.
This master equation is applicable for arbitrary coupling strengths and bias voltages.
In general, the nature of this master equation is non-Markovian,
as explicitly reflected by the time-dependent rates in \Eq{rates-wbl}.  \\
\\
{\it Markovian Limit}.---
As the usual treatment, one may consider the Markovian approximation
which is characterized by time independent decohrence/dissipative rates.
The Markovian approximation is largely associated with
taking a long time limit for the transient rates.
As is well known, in most cases, this treatment is very successful.
Now, in this work, let us consider the Markovian approximation
by taking the long time limit ($t\rightarrow\infty$)
for the transient rates given by \Eqs{rates-wbl}.
We obtain
\bsube\label{rates-steady}
\begin{align}
 \Gamma^{\pm}_{\alpha 11}& =\Gamma_{\alpha}
 \!\int\! d\omega\,
 n^{(\pm)}_{\alpha}(\omega) A_{11}(\omega),
  \label{rate11-steady}
 \\
  \Gamma^{\pm}_{\alpha 22}& =\Gamma_{\alpha}
 \!\int\! d\omega\,
 n^{(\pm)}_{\alpha}(\omega) A_{22}(\omega),
  \label{rate22-steady}
 \\
  \Gamma^{+}_{\alpha 12}&=- \Gamma^{-}_{\alpha 12}
  =\Gamma_{\alpha}
 \!\int\!d\omega\,
 n^{(+)}_{\alpha}(\omega) A_{12}(\omega).
 \label{rate12-steady}
\end{align}
\esube
Here, precisely following the Green's function theory,
we defined the spectral function through
\be\label{bmAw}
 {\bm A}(\omega)
=\frac{1}{\pi}{\rm Re} [{\bm u}(\omega)] \,,
\ee
while ${\bm u}(\omega)$ is the Fourier transformation
of the Green's function matrix ${\bm u}(t)$ in time domain,
i.e., ${\bm u}(\omega)=\int_{0}^{\infty}\!dt e^{i\omega t}{\bm u}(t)$.
Here the initial condition of ${\bm u}(t)$ has been considered,
which makes the Fourier and Laplace transformations identical.
From \Eq{ueq1}, performing a Laplace transformation,
we obtain the solution in frequency domain as
\begin{eqnarray}\label{u-solution}
{\bm u}(\omega)
=  \frac{i}{Z} \left(
\begin{array}{cc}
\omega+\varepsilon_{\M} + i\Gamma &  -i\Gamma_{\rm d}  \\
-i\Gamma_{\rm d} & \omega-\varepsilon_{\M} + i\Gamma  \\
\end{array}
\right) \,,
\end{eqnarray}
where
$Z=(\omega+\varepsilon_{\M}+i\Gamma)(\omega-\varepsilon_{\M}+i\Gamma)+\Gamma_{\rm d}^2$.
Then, the spectral functions are obtained as
\bsube\label{Aw}
\begin{align}
A_{11}(\omega) &=\frac{\Gamma\left[(\omega+\varepsilon_{\M})^2
+4\Gamma_{\rm L}\Gamma_{\rm R}\right]}{\pi|Z|^2},
\\
A_{22}(\omega) &=\frac{\Gamma\left[(\omega-\varepsilon_{\M})^2
+4\Gamma_{\rm L}\Gamma_{\rm R}\right]}{\pi|Z|^2},
\\
A_{12}(\omega)& =A_{21}(\omega) = \frac{ \Gamma_{\rm d}
\left( \omega^2-\varepsilon_{\M}^2
-4\Gamma_{\rm L}\Gamma_{\rm R}\right)}{\pi|Z|^2} \,.
\label{Aw12}
\end{align}
 \esube
In general, the spectral functions hold the properties of
$\!\int\!d\omega\,A_{11/22}(\omega)=1$
and $\!\int\!d\omega\,A_{12}(\omega)=0$.

\subsection{Time-Dependent Currents}

In master equation approach, the transport currents should be calculated
using the reduced density matrix $\rho(t)$.
This approach holds, very naturally, the advantage of
rendering the obtained currents time dependent.
One may notice that using other approaches,
such as the non-equilibrium Green's function approach
or the scattering matrix theory,
it is very inconvenient (or impossible)
to calculate the time-dependent transport currents.
Let us start to consider the average of the current operators $\hat I_{\alpha}$.
Using the Heisenberg equation, we have
$\hat I_{\alpha}=-e\frac{d}{dt}\hat{N}_{\alpha}
=i e[\hat{N}_{\alpha},H_{\T}]$,
where the total electron number operator of lead-$\alpha$
reads as $\hat{N}_{\alpha}=\sum_k c^\dg_{\alpha k}c_{\alpha k}$.
Then, we obtain
\begin{align}\label{curroper}
  \hat I_{\alpha}=-i e\big(Q^\dg_{\alpha}F_{\alpha}-F^\dg_{\alpha}Q_{\alpha}\big),
\end{align}
The current in lead-$\alpha$ is
$I_{\alpha}(t) \equiv\la \hat I_{\alpha}\big\ra_{\rm T}
={\rm Tr}_{\rm T}[\hat I_{\alpha}\rho_{\T}(t)]$,
where the average is over the total states of the whole system-plus-leads,
i.e., ${\rm Tr}_{\rm T}\equiv {\rm tr}_{\s}{\rm tr}_{\B}$.
Further, we reexpress the current in terms of the ADOs
\begin{align}\label{curr1}
 I_\alpha(t)&=-i e\,{\rm tr}_{\s}\big[Q^\dg_\alpha\rho^-_{\alpha }(t)
 -\rho^+_{\alpha }(t)Q_\alpha\big] \,.
 \end{align}
As mentioned above and demonstrated in \App{appendix}, the ADOs $\rho^{\pm}_{\alpha }(t)$
can be expressed in terms of the reduced density matrix $\rho(t)$, see \Eq{ADOsoper}.
Then, the current is obtained as
\begin{align}\label{curr}
  I_\alpha(t)&\!=\!
 I_{\alpha 11}(t)\!+\! I_{\alpha 22}(t)\!+\!(\!-\!)^\alpha I_{\alpha 12}(t)\!
 +\!(\!-\!)^\alpha I_{\alpha 21}(t) \,,
\end{align}
with the various component currents explicitly given by
\bsube\label{curr-rate}
\begin{align}
 I_{\alpha 11}(t)&= e
 \Big[\Gamma^{+}_{\alpha 11} (t)\rho_{00}(t)-\Gamma^{-}_{\alpha 11}(t)\rho_{11}(t)\Big],
 \label{I11}
\\
  I_{\alpha 22}(t)
 & =e
 \Big[\Gamma^{+}_{\alpha 22} (t)\rho_{11}(t)-\Gamma^{-}_{\alpha 22}(t)\rho_{00}(t)\Big],
 \label{I22}
\\
    I_{\alpha 12}(t)&=
e\left[\Gamma^{+}_{\alpha 12} (t)\rho_{00}(t)-
  \Gamma^{-}_{\alpha 12}(t)\rho_{11}(t)\right],
  \label{I12}
\\
  I_{\alpha 21}(t)&=
  e \left[\Gamma^{+}_{\alpha 21} (t)\rho_{11}(t)-
 \Gamma^{-}_{\alpha 21}(t)\rho_{00}(t)\right] \,.
 \label{I21}
\end{align}
\esube
In deriving this result, the elements of the density matrix were defined through
$\rho_{11}(t)={\rm tr}_{\s}[f^\dg f\rho(t)]$
and $\rho_{00}(t)={\rm tr}_{\s}[f f^\dg\rho(t)]$.
In practice, they should be solved from \Eq{EME-wbl}.
The various rates, $\Gamma^{\pm}_{\alpha ij}(t)$, are those given by \Eq{rates-wbl}.
Noting that $\Gamma^{\pm}_{\alpha 12} (t)=\Gamma^{\pm}_{\alpha 21} (t)$,
we thus have $ I_{\alpha 12}(t)= I_{\alpha 21}(t)$.

In terms of rate process, which applies also to understanding the above
master equation,
the physical meaning of the various current components can be understood as follow.
(i)
In the current $I_{\alpha 11}(t)$, the first term is associated with
normal tunneling process of an electron
from the $\alpha$-lead to the MZMs quasiparticle,
while the second term describes the inverse process of the first one.
(ii)
In the current $I_{\alpha 22}(t)$,
the first term describes the Andreev reflection (AR) process
of an electron entering the MZMs from the $\alpha$-lead
and annihilating the existing quasiparticle,
associated with formation of a Cooper pair;
and the second term describes the inverse process of the first one,
associated with splitting a Cooper pair.
(iii)
In $I_{\alpha 12}(t)$,
the first term describes another contribution
of normal tunneling process
of an electron from the $\alpha$-lead to the MZMs quasiparticle,
while the second term is the inverse process of the first one.
(iv)
In $I_{\alpha 21}(t)$, again,
the first and second terms describe additional contribution of the
AR process and its inverse process,
associated with formation and splitting of a Cooper pair, respectively.

Importantly,
all the rates in $I_{\alpha 11}(t)$ and $I_{\alpha 22}(t)$,
i.e., $\Gamma^{\pm}_{\alpha 11}$ and $\Gamma^{\pm}_{\alpha 22}$,
involves internal self-energy propagation associated with
the {\it normal} Green's functions $u_{11}(t,\tau)$ and $u_{22}(t,\tau)$.
By contrast,
all the rates in $I_{\alpha 12}(t)$ and $I_{\alpha 21}(t)$,
i.e., $\Gamma^{\pm}_{\alpha 12}$ and $\Gamma^{\pm}_{\alpha 21}$,
involves internal self-energy propagation associated with
the {\it anomalous} Green's functions $u_{12}(t,\tau)$ and $u_{21}(t,\tau)$.

\section{Numerical Results and Limiting Case Analysis}

\subsection{Transient Rates and Currents}

The above formal results of master equation and currents
can be applied to transport under arbitrary bias voltages.
For the two-lead (three-terminal) transport setup,
following Ref.\,\onlinecite{Nil08120403},
we simply consider the equally biased voltage of the leads
with respect to the chemical potential
of the grounded superconductor (which is set as zero),
i.e., $\mu_{\rm L}/e=\mu_{\rm R}/e=V$.
For this bias setup, the currents flow back to the leads
from the superconductor through the grounded terminal.
Between the leads, there is no transmission current from one lead to another.

In Fig.\ 2(a), we illustrate the transient behavior of the rates.
For simplicity, in the numerical studies of this work,
we assume the wide-band approximation for the leads.
Rather than the usual constant rates, the transient behavior of the rates
--even under the wide-band limit--reflects the non-Markovian nature.
Using the non-Markovian transient rates,
we solve the master equation and calculate the transient current,
with the results as shown in Fig.\ 2(b) and (c) by the solid curves.
As an interesting comparison, we also use the asymptotic rates
$\Gamma^{\pm}_{\alpha ij}(t\to\infty)$
to solve the master equation and calculate the transient current,
obtaining the results as shown in Fig.\ 2(b) and (c) by the dashed curves.
From Fig.\ 2(b), we find that the occupation probability does not
differentiate from each other too much.
However, in Fig.\ 2(c), we find that
the total current $I_{\rm L}(t)$ from the Markovian rates
does not reveal any transient behavior.
A simple explanation for this result is as follows.
The total current $I_{\rm L}(t)$ contains two parts:
one is from the normal tunneling process, which associates with
electron entering the superconductor conditioned on the MZMs unoccupied;
another part is from the Andreev reflection process,
which is conditioned on the MZMs occupied.
Then, the total current becomes free from the occupation of the MZMs.

In Fig.\ 2(e)-(f), we show the transient behavior of the component currents.
As expected, even using the asymptotic rates,
the switching-on transient behavior of $I_{{\rm L}11}(t)$ and $I_{{\rm L}22}(t)$ is obvious,
in contrast to the total current $I_{\rm L}(t)$.
We notice that, at the very beginning stage, $I_{{\rm L}22}(t)$ is negative.
This is because, under the finite bias voltage $V$ (not large enough),
the empty-occupation of the MZMs allows for happening of the inverse AR process,
which splits a Cooper pair and results in occupation of the MZMs
and another electron entering the lead from the superconductor,
causing thus the negative $I_{{\rm L}22}(t)$ as observed.
In Fig.\ 2(f), we find that the {\it small} off-diagonal current components are negative.
This can be understood using the negative and very small
off-diagonal rates shown in Fig.\ 2(a).
Actually, as to be further analyzed in the following,
at large $V$ limit, the off-diagonal rates will vanish.
Thus, the off-diagonal currents will be zero under such limit.
Finally, we remark that the results displayed by the solid curves in Fig.\ 2(e)-(f)
are from using the non-Markovian transient rates as shown in Fig.\ 2(a).
The more complicated transient behaviors are owing to
the gradual formation of the asymptotic rates, during the rate process dynamics.

\begin{figure}
\centering
\centerline{\includegraphics*[width=1.\columnwidth,angle=0]{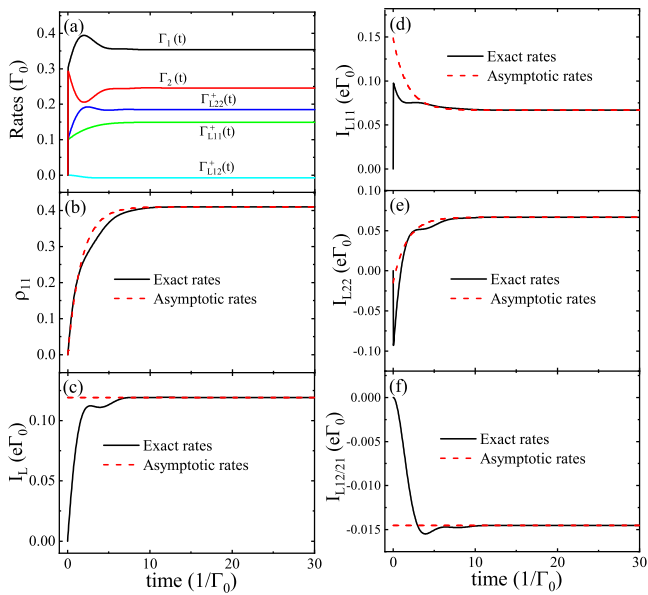}}
 \caption{ (Color online)
Numerical results for the transient rates, occupation probability, and currents.
All energies are measured in (arbitrary) unit of $\Gamma_0$,
and the unit of time is $1/\Gamma_0$.
Parameters assumed in the simulation are:
$\mu_{\rm L}=\mu_{\rm R}=0.8 \Gamma_{0}$
$\Gamma_{\rm L}=2\Gamma_{\rm R}=0.2 \Gamma_{0}$,
$\varepsilon_{\M}=0.5\Gamma_0$, and $k_{\rm B}T=0.05 \Gamma_{0}$.
The transient rates and occupation probability
are shown in (a) and (b);
the total left-lead current and the various component currents
are shown in (c) and (d)-(f), respectively.
For the occupation probability and currents,
results from using the exact transient rates (black-solid curves)
and using the asymptotic rates (red-dashed curves),
are plotted against each other for comparison.
  \label{fig2}}
\end{figure}
%


 \begin{figure}
\centering
\centerline{\includegraphics*[width=1.\columnwidth,angle=0]{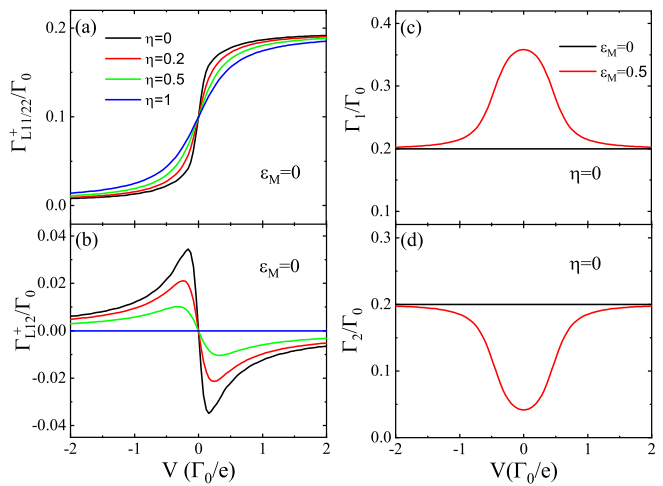}}
 \caption{ (Color online)
Long-time-limit asymptotic rates
as a function of the bias voltage $V$,
under also the symmetric bias setup $\mu_{\rm L}/e=\mu_{\rm R}/e=V$
with respect to the chemical potential of the superconductor.
The coupling asymmetry parameter $\eta=\Gamma_{\rm L}/\Gamma_{\rm R}$
and the Majorana coupling energy $\varepsilon_M$
are altered as shown in the figure.
The component rates are plotted in (a) and (b),
while the total rates $\Gamma_1$ and $\Gamma_2$ are shown in (c) and (d).
Parameters used are the same as in \Fig{fig2}.
  \label{fig3}}
\end{figure}

\subsection{Markovian Limit}

Markovian approximation can considerably simplify the master equation approach,
making it extremely convenient for studying, not only the transport currents,
but also current correlations and full-counting statistics.
In this subsection, we analyze the conditions for Markovian approximation to the MME.

As briefly shown in Sec.\ II A, the Markov approximation is largely associated with
taking a long-time limit for the transient rates,
in present work, which are given by \Eqs{rates-wbl}.
Hereafter, let us term the long-time asymptotic rates as Markovian rates.
In Fig.\ 3, we display the behaviors of the Markovian rates,
via their dependence on the bias voltage
and the coupling asymmetry parameter $\eta=\Gamma_{\rm L}/\Gamma_R$.
Owing to the a few symmetry properties of the rates $\Gamma^{\pm}_{\alpha ij}$,
here we only plot the results of $\Gamma^+_{{\rm L}11/22}$ and $\Gamma^+_{{\rm L}12}$.

In Fig.\ 3(a), for the diagonal rates $\Gamma^+_{{\rm L}11/22}$, we find that
they increase with the bias voltage $V$ (from negative to positive),
while at the large $\pm V$ limits they approach, respectively, saturated values and zero.
This bias-voltage dependence can be understood based on the lead-electron-occupation,
which determines the integrated range of contribution to the rates.
In Fig.\ 3(b), as an example of the off-diagonal rates,
we find an antisymmetric dependence on $\pm V$.
This behavior is determined by the following properties of the
off-diagonal element of the spectral function matrix from \Eq{Aw12},
i.e., $A_{12}(\omega)=A_{12}(-\omega)$ and
$\int^{\infty}_{-\infty}d\omega A_{12}(\omega) =0$.
Using them, we can easily prove that
$\Gamma^+_{{\rm L}12}(V)+ \Gamma^+_{{\rm L}12}(-V) =0 $.
In Fig.\ 3(c) and (d), we show the total rates $\Gamma_1$ and $\Gamma_2$
which characterize, respectively,
the annihilation and creation rates of the MZMs-quasiparticle.
For $\varepsilon_{\rm M}=0$, we find that the both total rates are bias ($V$) independent;
while for $\varepsilon_{\rm M}\neq 0$, they reveal a symmetric dependence on $\pm V$
and take maximum and minimum at $V=0$ for $\Gamma_1$ and $\Gamma_2$, respectively.
The reason for this behavior is as follows.
First, based on \Eq{Aw12}, we know $A_{12}(\omega)=A_{21}(\omega)$.
Then, we can prove that $\Gamma^-_{{\rm L}12}(V)+ \Gamma^+_{{\rm L}12}(V)
= \Gamma_{\rm L} \int^{\infty}_{-\infty}d\omega A_{12}(\omega) =0 $.
Second, for $\varepsilon_M\neq 0$, we know that
the peaks of $A_{11/22}(\omega)$ are centered at $\pm \varepsilon_M$.
Then, we know that $\Gamma^+_{{\rm L}11}(V)$ increases with $V$
later than $\Gamma^+_{{\rm L}22}(V)$,
when crossing the region around zero bias (from negative to positive).
Because of the same reason, $\Gamma^-_{{\rm L}11}(V)$ decreases with $V$
earlier than $\Gamma^-_{{\rm L}22}(V)$, when crossing the region around zero bias.
As a result, the total rates,
$\Gamma_1(V)=\Gamma^-_{{\rm L}11}(V)+\Gamma^+_{{\rm L}22}(V)$
and $\Gamma_2(V)=\Gamma^+_{{\rm L}11}(V)+\Gamma^-_{{\rm L}22}(V)$,
exhibit the behavior as shown in Fig.\ 3(c) and (d).

\begin{figure}
\centering
\centerline{\includegraphics*[width=1.\columnwidth,angle=0]{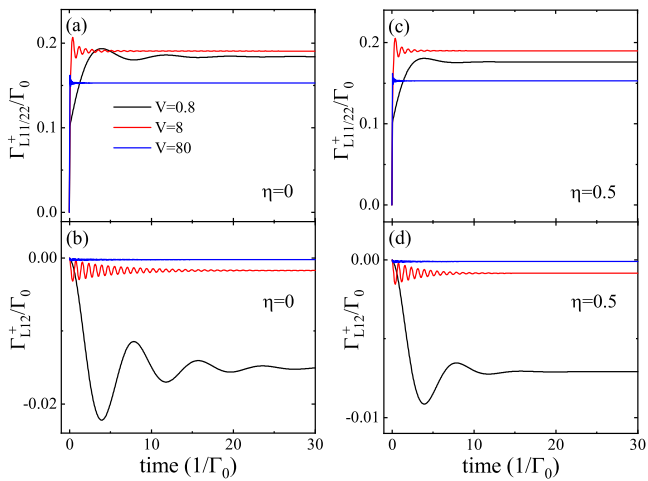}}
 \caption{ (Color online)
Transient behaviors of the component rates under different bias voltages,
indicating the large-bias condition for Markovian approximation.
Results for the single-lead setup ($\eta=0$) are plotted in (a) and (b),
while for the two-lead setup ($\eta=\Gamma_{\rm L}/\Gamma_{\rm R}=0.5$) in (c) and (d).
Parameters used in this plot are the same as in \Fig{fig2}.
  \label{fig4}}
\end{figure}

In Fig.\ 4 we show further
the transient behaviors of the rates under different bias voltages.
We see that for small bias voltage,
the transient behavior (non-Markovian feature) is remarkable,
while with increase of the bias voltage
the rates more rapidly approach the (Markovian) asymptotic results.
In practice, when the bias voltage is considerably larger than the broadening width,
i.e., in the sequential tunneling regime,
the Markovian master equation approach should work very well.
In Fig.\ 4(a) and (b), we show the transient behaviors of diagonal and off-diagonal rates
for single-lead coupling, while in (c) and (d) for a two-lead coupling configuration.
We find qualitatively similar behaviors for both coupling configurations.
This is because the rates $\Gamma^{\pm}_{\alpha ij}$
are dominantly affected by the Fermi occupation of the $\alpha$-lead electrons,
while influence of the opposite lead is weakly mediated through
the propagating function ${\bm u}(\omega)$ via the self-energy effect.

Finally, we carry out the analytic results under wide-band and large-bias limits.
At large $V$ limit, $\mu_{\alpha}\to \infty$,
we have $n^+_{\alpha}=1$ and $n^-_{\alpha}=0$.
Then, straightforwardly, one can prove that
$g^{+}_{\alpha}(t-\tau)\to \Gamma_{\alpha}\delta(t-\tau)$,
which yields $\Gamma^+_{\alpha 11} = \Gamma^+_{\alpha 22} = \Gamma_{\alpha}$.
Owing to $n^-_{\alpha}=0$, we simply have $\Gamma^{-}_{\alpha ij}=0$.
Substituting these results into \Eq{EME-wbl},
we obtain a very simple Lindblad-type master equation
which, however, contains the central physics of Andreev process
associated with the MZMs, and makes the results different from
transport through the conventional single-level quantum dot.

\subsection{Steady-State Currents: Connection with Other Approaches}

Let us introduce the more conventional (retarded) superconductor Green's functions
\bsube 
\begin{align}
G^r_{11}(t,\tau) &\!=\! -i\Theta(t-\tau)\la\{f(t), f^{\dagger}(\tau)\}\ra \,,
\\
G^r_{22}(t,\tau) &\!=\! -i\Theta(t-\tau)\la\{f^{\dagger}(t), f(\tau)\}\ra \,,
\\
G^r_{12}(t,\tau) &\!=\! -i\Theta(t-\tau)\la\{f(t), f(\tau)\}\ra   \,,
\\
G^r_{21}(t,\tau) &\!=\! -i\Theta(t-\tau)\la\{f^{\dagger}(t), f^{\dagger}(\tau)\}\ra  \,.
\end{align}
\esube
From the equation-of-motion of ${\bm u}(t,\tau)$, i.e., Eq(..),
we know that ${\bm u}(t,\tau)=i{\bm G}^r(t,\tau)$.
Then, the spectral function matrix can expressed as
\bea
{\bm A}(\omega) = \frac{1}{\pi} {\rm Re}[{\bm u}(\omega)]
= -\frac{1}{\pi} {\rm Im}[{\bm G}^r(\omega)]
\eea
From the Dyson equation, we have
\bea\label{dyson-1}
&& {\rm Im}[{\bm G}^r(\omega)] = \frac{1}{2i}[{\bm G}^r(\omega)-{\bm G}^a(\omega)]  \nl
&&  ~~~~ = -i [ {\bm G}^r(\omega) \wti{\bm \Sigma}^r(\omega) {\bm G}^a(\omega)  ]
\eea
The retarded self-energy matrix is given by
\bea
\wti{\bm \Sigma}^r(\omega) = -\frac{i}{2}
\sum_{\alpha={\rm L,R}}
\left[{\bm\Gamma_{\alpha}} (\omega) +  {\bm\Gamma_{\alpha}} (-\omega) \right] \,,
\eea
while the rate matrices read as
\bsube\label{Gamm-mat}
\begin{align}
{\bm \Gamma}_{\rm L} (\omega)
&={\Gamma}_{\rm L} (\omega)\left(
                                    \begin{array}{cc}
                                      1 & 1 \\
                                      1 &  1 \\
                                    \end{array}
                                  \right),
\\
 {\bm \Gamma}_{\rm R} (\omega)
 &={\Gamma}_{\rm R} (\omega)\left(
 \begin{array}{cc}
  1 &  -1 \\
    -1&  1\\
   \end{array}
   \right).
\end{align}
\esube
As to be clear in the following,
we may denote
$ {\bm \Gamma}_{\alpha} (\omega)
 \equiv {\bm \Gamma}^e_{\alpha}$
and $ {\bm \Gamma}_{\alpha} (-\omega)\equiv {\bm \Gamma}^h_{\alpha}$.

Based on \Eq{curr-rate}, i.e.,
the time-dependent current formula of the master equation approach,
the non-equilibrium Green's function
results of steady-state currents can be easily obtained.
As an example, let us consider the steady-state current
$I_{\rm L11}=\Gamma^+_{\rm L11}-\Gamma_{\rm L} \bar{\rho}_{11}$.
We have
\bea
\Gamma^+_{{\rm L}11} &=& -2\Gamma_{\rm L} \int \frac{d\omega}{2\pi}
n^+_{\rm L} (\omega) {\rm Im}[{\bm G}^r(\omega)]_{11}   \,, \nl
\bar{\rho}_{11} &=& \int \frac{d\omega}{2\pi}{\rm Im}[G_{11}^<(\omega)]   \,.
\eea
 Here we have applied the relationship between
$\rho_{11}(t) = {\rm tr}_{\rm s} [f^{\dagger}f \rho(t)]$
and $G_{11}^<(t,\tau)=i\la f^{\dagger}(\tau)f(t) \ra$.
After similar treatment to all the other component currents
and applying \Eq{dyson-1} and the well known Keldysh equation for ${\bm G}^<(\omega)$,
combination (together also
with a procedure of symmetrization)
of the component currents yields
 \begin{align}\label{currs1}
 I _{\rm L}&=\frac{e}{2h}\int \!d\omega
  \big\{{\cal T}^{ee}_{\rm LR}(\omega)
   \big[n^+_{\rm L} (\omega)- n^+_{\rm R} (\omega)\big]
  \nl&\quad\quad\quad\quad
   +  {\cal T}^{hh}_{\rm LR}(\omega)
   \big[n^-_{\rm R}  (-\omega)- n^-_{\rm L}  (-\omega)\big]\big\}
 \nl&
 +\frac{e}{2h}\int \!d\omega
   \big[ {\cal T}^{eh}_{\rm LLA}(\omega)+{\cal T}^{he}_{\rm LLA}(\omega)\big]
   \big[n^+_{\rm L} (\omega)- n^-_{\rm L}  (-\omega)\big]
 \nl&+
 \frac{e}{2h}\int \!d\omega
   \big\{{\cal T}^{eh}_{\rm LR A}(\omega)
   \big[n^+_{\rm L} (\omega)- n^-_{\rm R}  (-\omega)\big]
   \nl&\quad\quad\quad\quad
   +  {\cal T}^{he}_{\rm LRA}(\omega)
   \big[ n^+_{\rm R} (\omega)- n^-_{\rm L}  (-\omega)\big]\big\}  \,.
  \end{align}
In this result, the various transmission coefficients are given by
\bea\label{T-LR}
{\cal T}^{eh}_{\rm L L A}(\omega)
 &=& {\rm Tr}\big[{\bm \Gamma}^e_{\rm L} {\bm G}^r (\omega)
 {\bm \Gamma}^h_{\rm L}   {\bm G}^a (\omega)\big] \,,  \nl
 {\cal T}^{he}_{\rm L L A}(\omega)
 &=& {\rm Tr}\big[{\bm \Gamma}^h_{\rm L} {\bm G}^r (\omega)
 {\bm \Gamma}^e_{\rm L}   {\bm G}^a (\omega)\big] \,, \nl
{\cal T}^{ee}_{\rm LR}(\omega) &=&
 {\rm Tr}\big[{\bm \Gamma}^e_{\rm L} {\bm G}^r (\omega)
 {\bm \Gamma}^e_{\rm R}   {\bm G}^a (\omega)\big] \,,  \nl
 {\cal T}^{hh}_{\rm LR}(\omega)&=&
 {\rm Tr}\big[{\bm \Gamma}^h_{\rm L} {\bm G}^r (\omega)
 {\bm \Gamma}^h_{\rm R}   {\bm G}^a (\omega)\big] \,,  \nl
{\cal T}^{eh}_{\rm LRA}(\omega) &=&
 {\rm Tr}\big[{\bm \Gamma}^e_{\rm L} {\bm G}^r (\omega)
 {\bm \Gamma}^h_{\rm R}   {\bm G}^a (\omega)\big] \,,  \nl
{\cal T}^{he}_{\rm LRA}(\omega) &=&
 {\rm Tr}\big[{\bm \Gamma}^h_{\rm L} {\bm G}^r (\omega)
 {\bm \Gamma}^e_{\rm R}   {\bm G}^a (\omega)\big] \,.
\eea
The meaning of these transmission coefficients is clear:
${\cal T}^{eh(he)}_{\rm L L A}$
describes the local Andreev-reflection process
associated with incidence of an electron (a hole)
and reflection of a hole (an electron) in the same (left) lead;
${\cal T}^{ee(hh)}_{\rm L R}$ describes
the normal transport process
of an electron (or a hole) from the left to the right lead;
and ${\cal T}^{eh (he)}_{\rm L R A}$ describes
the crossed Andreev-reflection process
with incidence of an electron (a hole) from the left lead
and reflection of a hole (an electron) in the right lead.

We may remark that, when reaching the compact form of the above results,
we have properly reorganized the individual contributions
in the various component currents.
For instance, based on $I_{\rm L}=\sum_{i,j=1,2} I_{{\rm L} ij}$,
we have performed the following combination
of elements from the four component currents
\bea\label{T-sum}
&&\sum_{i,j=1,2} \Gamma^e_{\rm L} ({\bm G}^r{\bm \Gamma}^e_{\rm R}{\bm G}^a )_{ij} \nl
&& = \Gamma^e_{\rm L} \left( M^e_{{\rm R}11} +M^e_{{\rm R}21}
  + M^e_{{\rm R}12} + M^e_{{\rm R}22} \right)  \nl
&& = {\rm Tr}({\bm \Gamma}^e_{\rm L} {\bm M}^e_{\rm R})
\eea
Here we introduced the $2\times 2$ matrix
${\bm M}^e_{\rm R}={\bm G}^r{\bm \Gamma}^e_{\rm R}{\bm G}^a$.
It is also worth noting that
the individual elements $\Gamma^e_{\rm L} M^e_{{\rm R}ij} $ in the above combination
are from the rate components $\Gamma^{\pm}_{{\rm L}ij}$ in the master equation,
i.e., $\Gamma^{\pm}_{{\rm L}ij}
=\Gamma^e_{\rm L} \int d\omega n^{\pm}_{{\rm L}}(\omega) A_{ij}(\omega)$.

Finally, we remark that the electron- and hole-type rates,
${\bm \Gamma}^{e/h}_{\alpha}(\pm\omega)$,
are originated from the BdG-type consideration for the original electrons in the leads.
Owing to the particle-hole symmetry,
the energy replacement of $\omega\to -\omega$ should be made
when considering the transformation from an electron to a hole.
Therefore, from the original rates, we redefined
$ {\bm \Gamma}_{\alpha} (\omega)={\bm \Gamma}^e_{\alpha}$
and $ {\bm \Gamma}_{\alpha} (-\omega)= {\bm \Gamma}^h_{\alpha}$.
Moreover, associated with this conversion,
the electron and hole occupation function are different,
i.e., the former is $n^+_{\alpha}(\omega)$
and the latter is $n^-_{\alpha}(-\omega)$, as one can find in \Eq{currs1}.

In this context, based on \Eq{T-LR} and in particular
its construction as exemplified by \Eq{T-sum}
(i.e., the combination of multiple channels),
we briefly discuss the issue of teleportation, in an attempt to shed light on
the reason of the vanishing of the teleportation channel when $\varepsilon_{\M}\to 0$.
To see this more clearly, let us reexpress the rate matrices in \Eq{Gamm-mat}
in terms of the projection operators as
\bea\label{Gamma-mat-2}
{\bm \Gamma}_{\rm L} &=& 2\Gamma_{\rm L} |\Phi_{\rm L}\ra \la\Phi_{\rm L}|  \,,  \nl
{\bm \Gamma}_{\rm R} &=& 2\Gamma_{\rm R} |\Phi_{\rm R}\ra \la\Phi_{\rm R}|  \,,
\eea
where $|\Phi_{\rm L/R}\ra=(|1\ra \pm |2\ra)/\sqrt{2}$,
with $|1\ra$ and $|2\ra$ the basis states
which support the expression of the matrices in \Eq{Gamm-mat}.
In terms of the BdG formalism language,
they correspond to the positive and negative energy states
$|E_0\ra$ and $|-E_0\ra$, respectively, with $E_0=\varepsilon_{\M}$
the energy of the regular fermion (the $f$-particle) associated with the MZMs.
Accordingly, the states $|\Phi_{\rm L}\ra$ and $\Phi_{\rm R}\ra$
correspond to the left and right Majorana modes.

Based on \Eq{Gamma-mat-2},
one can examine that
${\bm \Gamma}_{\rm L} {\bm G}^r (\omega){\bm \Gamma}_{\rm R} \to 0$
at the limit $\varepsilon_{\M}\to 0$, which is valid
for all the $ee$, $hh$, $eh$ and $he$ transmission components.
Actually, using the explicit solution of \Eq{u-solution}, we have
$\la\Phi_{\rm L} | {\bm u} |\Phi_{\rm R}\ra =
u_{11}-u_{22}-u_{12} + u_{21}=2\,i\,\varepsilon_{\M}/ Z$,
while noting that ${\bm G}^r =-i{\bm u}$.
This is the remarkable issue of teleportation vanishing
when the Majorana coupling energy approaches to zero,
which implies that all the electron transmission and the crossed Andreev-reflection process
through a pair of MZMs vanish at the limit $\varepsilon_{\M}\to 0$.

Here, we would like to point out that this result is a consequence of
the combination of multiple transmission channels.
It does not involve only a single particle transmission,
but involves several different charge configurations.
For single transmission process,
the vanishing of teleportation
does not take place when $\varepsilon_{\M}\to 0$.
For instance, let us look at
\bea
&&\Gamma^e_{\rm L}  M^e_{{\rm R}11}
= \Gamma^e_{\rm L}
\la 1| {\bm G}^r{\bm \Gamma}^e_{\rm R}{\bm G}^a  |1\ra   \nl
&&= 2\Gamma^e_{\rm L} \Gamma^e_{\rm R}
\la 1| {\bm G}^r|\Phi_{\rm R}\ra \la\Phi_{\rm R}|{\bm G}^a  |1\ra  \,.
\eea
It does not vanish as $\varepsilon_{\M}\to 0$.
Therefor,
the result that an electron cannot transmit through a pair of MZMs
at the limit $\varepsilon_{\M}\to 0$ is an effective picture
owing to combination of multiple processes.
Under large bias condition (i.e., with the bias voltage
larger than the zero-energy-level broadening),
the multiple processes cannot take place simultaneously.
Then, one should not expect the vanishing phenomenon of teleportation.

The result of transmission coefficients given by \Eq{T-LR} can be connected
with the stationary $S$-matrix scattering approach as follows.
Let us introduce the $S$-matrix (operator) as
\bea
{\bm S}_{\rm LR}(\omega) = {\bm W}_{\rm L}
{\bm G}^r(\omega) {\bm W}^{\dagger}_{\rm R}  \,,
\eea
where the two coupling operators
between the left/right Majorana modes and the left/right leads
are give by
\bea
{\bm W}_{\rm L} &=& \sum_{p=e,h}\sqrt{\Gamma_{\rm L}}
\, |p_{\rm L}\ra \la\Phi_{\rm L} |   \,,  \nl
{\bm W}^{\dagger}_{\rm R} &=& \sum_{q=e,h}\sqrt{\Gamma_{\rm R}}
\, |\Phi_{\rm R}\ra \la q_{\rm R}|   \,.
\eea
Here, $p$ and $q$ are introduced to denote the electron and hole states
in the left and right leads, respectively.
Then, the coefficients of transmission from the left to right leads
can be reexpressed in terms of the $S$-matrix elements as
\bea
{\cal T}^{pq}_{\rm LR}(\omega)
= | \la p_{\rm L}|{\bm S}_{\rm LR}(\omega)|q_{\rm R}\ra |^2  \,.
\eea
All the other coefficients (for the local Andreev reflection) in \Eq{T-LR}
can be similarly formulated by means of the $S$-matrix scattering approach.

\section{Summary and Discussion}

We have presented a master equation approach for transport through MZMs.
The master equation approach will be convenient for studying the various
statistical properties of transport currents
and important time-dependent phenomena
such as photon-assisted tunneling through the MZMs
and modulation effect of the Majorana coupling energy.
In this initial work,
we carried out explicit results
for the master equation and currents.
We illustrated the behaviors of transient rates, occupation dynamics and currents
and exploited the Markovian condition for the rates,
which can extremely simplify the applications in practice.
Via careful analysis for the structure of the rates,
we also revealed the intrinsic connection between
the master equation and the BdG $S$-matrix scattering approach,
by noting that the charge transfer picture and occupation dynamics
in both formulations are quite different.
This connection is absent in literature
and can help us to better understand the issue of Majorana teleportation.

About the possible application to time-depend transports,
the simplest case is to consider modulating the central system
under Markovian approximation for the leads.
In this case, in the master equation, only the central system Hamiltonian
in the coherent term becomes time dependent, and the rates in the dissipative terms
are unaffected by the time-dependent modulation of the central system.
Beyond Markovian approximation, more rigorous treatment for the rates
needs to consider the effect of modulation on
the reduced propagating function ${\bm u}(t,\tau)$,
for the central system of MZMs, whose equation-of-motion is given by \Eq{ueq1}.
If we consider a modulation to the coupling energy between the MZMs,
we only need to insert the time-dependent $\varepsilon_{\M}(t)$ into \Eq{ueq1}
to solve ${\bm u}(t,\tau)$ and calculate the time-dependent rates.
Then, solving the master equation allows us to obtain the time-dependent currents.

An important example of time-dependent transport is to consider adding an ac voltage to the dc bias.
In this case, rather than treating time-dependent chemical potentials for the leads,
one can alternatively keep the occupation of the lead electrons (labelled by ``$\alpha k$")
unchanged (determined by the dc bias),
but at the same time let the energy levels of the lead electrons be time dependent,
i.e., $\varepsilon_{\alpha k}(t)=\varepsilon_{\alpha k}+\Delta_{\alpha}(t)$.
In this treatment, we only need to perform a replacement
for the correlation functions of the lead electrons
$(g_{\alpha}, g^{\pm}_{\alpha})\to (\tilde{g}_{\alpha}, \tilde{g}^{\pm}_{\alpha})$,
while the latter read as
\bsube\label{corr-bath-time}
 \begin{align}
    \tilde{g}_{\alpha}(t,\tau)
 &=\exp\left[-i\,\int_{\tau}^{t}\! \Delta_{\alpha}(\tau')d\tau'\right] g_{\alpha}(t-\tau)
 \label{corr-b1} ,
 \\
    \tilde{g}^{\pm}_{\alpha}(t,\tau)
 &=
 \exp\left[-i\,\int_{\tau}^{t}\! \Delta_{\alpha}(\tau')d\tau'\right]g^{\pm}_{\alpha}(t-\tau),
 \end{align}
 \esube
where $g_{\alpha}(t-\tau)$ and $g^{\pm}_{\alpha}(t-\tau)$ are given by \Eq{corr-bath}.
Inserting these extended functions into the expressions of the rates,
one can straightforwardly solve the master equation and calculate the time-dependent currents.
Obviously, this formulation works for arbitrary time-dependent voltages,
such as step-like modulation and rectangular bias pulse.

For transport through MZMs,
existing studies were largely restricted in steady state investigations,
either for a one-lead or two-lead transport setup.
When considering an ac bias applied to the dc voltage, the interesting problem
of photon-assisted tunneling through the MZMs can be analyzed in detail,
by applying the master equation approach.
One may expect that this type of investigations
can provide additional dynamical insight
beyond the zero-bias-peak of conductance.
One may also expect some profound features which can distinguish
the photon-assisted tunneling through the MZMs
from through the conventional resonant-tunneling diodes,
by noting that the latter has been an important subject
and received considerable attentions \cite{Hau08}.

\vspace{0.7cm}
{\flushleft\bf
ACKNOWLEDGMENTS.}---
This work was supported by the National Key Research
and Development Program of China (Grant No.\ 2017YFA0303304)
and the NNSF of China (Grant No.\ 11974011).

\appendix

\section{Derivation of the Master Equation and Current Formula}
\label{appendix}

Starting with the Liouvillian equation for the entire state $\rho_{\T}(t)$,
$\dot{\rho}_{\T}(t)=-i[H_{\T},\rho_{\T}(t)]$, we obtain
\begin{align}\label{LvEq}
\dot{\rho}(t)&\!\!=\!\!-i[H_{\s},\rho(t)]\!-\!i\!\sum_{\alpha}[Q_\alpha,\rho^{+}_{\alpha}\!(t)]
\!-\!i\!\sum_{\alpha}[Q^\dg_\alpha,\rho^{-}_{\alpha}\!(t)],
\end{align}
where $\rho(t)={\rm tr}_{\B}(\rho_{\T}(t))$ is the reduced density matrix,
and the other two auxiliary density operators (ADOs) are introduced as
$\rho^+_{\alpha }(t)\!\equiv\!\mb{tr}_{\B}\{\rho_{\T}(t)F^\dg_\alpha \}$
and $\rho^-_{\alpha }(t)\!\equiv\!\mb{tr}_{\B}\{F_\alpha \rho_{\T}(t)\}$.
In this Appendix, following the treatment developed in
previous work \cite{Jin10083013}, 
we establish a relation between $\rho^{\pm}_{\alpha }(t)$ and $\rho(t)$.
The basic idea is to establish
the propagation of $\rho(t)$ and $\rho^{\pm}_{\alpha }(t)$, respectively,
in terms of path integral by means of the fermionic
coherent state representation.
Then, removing the representation basis,
we extract the operator form for the relation between
$\rho^{\pm}_{\alpha }(t)$ and $\rho(t)$.

\subsection{Path-integral formulation for the propagation \\
of the reduced density matrix $\rho(t)$}

Within the path-integral formulation
using the fermionic coherent state representation,
the propagation of the reduced density matrix $\rho(t)$
can be expressed as  \cite{Tu08235311,Jin10083013}
\begin{align}\label{rdm}
\langle\xi_f|\rho(t)|\xi'_f\rangle=\int d\mu(\xi_0)& d\mu(\xi'_0)
\langle\xi_0|\rho(t_0)|\xi'_0\rangle \nonumber \\ & \times {\cal
J}(\xi^\ast_f,\xi'_f,t|\xi_0,\xi'^{\ast}_0,t_0)  \,.
\end{align}
The variables $\xi^\ast_f,\xi'_f,\xi_0,\xi'^{\ast}_0$ are the Grassmannian numbers
introduced through the fermionic coherent states:
 $f|\xi\rangle
 = \xi  |\xi \rangle$ and $\langle \xi |f^\dag =  \langle \xi |
 \xi^* $.
 They obey $\int d\mu(\xi)|\xi\ra\la\xi|=1$ with
 the integration measure defined by $d\mu(\xi)=d\xi^\ast d\xi e^{-{|\xi|^2}}$.
The propagating function is given by
\begin{align} \label{ppgf}
 {\cal J}(\xi^{\ast}_f,\xi'_f,t|\xi_0, \xi'^{\ast}_0,t_0)\!=\!\!
 \int\!\!\mathcal{D}[ \bm\xi ; \bm\xi' ]
 e^{iS_{\rm eff}[ \bm\xi;\bm\xi']}
\mathcal{F}
[ \bm\xi;\bm\xi'],
\end{align}
where $\bm\xi\equiv (\xi^{\ast}\!,\xi)$, and
$S_{\rm eff}[ \bm\xi;\bm\xi']=S_{\M}[ \bm\xi]-S^*_{\M}[ \bm\xi']$ is
the free action of the MZMs, with
$S_{\M}[  {\xi^\ast} , {\xi }]
=\big\{
-i  {\xi^\ast} \xi ( t )
+\int_{0}^{t}d\tau \big[i
 {\xi}^\ast(\tau)\dot{\xi}(\tau)
-H_{\M}(  {\xi^\ast}(\tau) , \xi(\tau))\big]\big\}$.
$\mathcal{F}[  \bm\xi;,\bm\xi']$ is the Feynman-Vernon-influence-functional,
after integrating out the degrees of freedom of the reservoirs (lead electrons),
which reads
\begin{align}\label{IF1}
{\cal F}[\bm\xi;\bm\xi']
 &=\exp\Big\{\!\!
-\!\!\int^t_{t_0}\!d\tau\int^{\tau}_{t_0}d\tau'
\hat{\bm \xi}^\dg(\tau)
\,{\bm g}(\tau,\tau')\,
\hat{\bm \xi}(\tau')
\nl&\quad\quad\quad
\!-\! \int^t_{t_0}\!d\tau\int^{\tau}_{t_0}\!d\tau'
\hat{\bm \xi}'^\dg(\tau)
\,{\bm g}(\tau',\tau)\,
\hat{\bm \xi}'(\tau')
\nl&\quad\quad\quad
 \!-\!\int^t_{t_0}\!d\tau\int^{t}_{t_0}\!d\tau'
\hat{\bm \xi}'^\dg(\tau)
\,{\bm g}(\tau,\tau')\,
\hat{\bm \xi}(\tau')
\nl&\quad\quad\quad
 \!+\!\int^t_{t_0}d\tau\int^{t}_{t_0}d\tau'
\hat{\bm \xi}^\dg(\tau)
\,{\bm {\wti g}}(\tau,\tau')\,
\hat{\bm \xi}(\tau')
\nl&\quad\quad\quad
\! +\!\int^t_{t_0}\!d\tau\int^{t}_{t_0}\!d\tau'
\hat{\bm \xi}^\dg(\tau)
\,{\bm {\wti g}}(\tau,\tau')\,
\hat{\bm \xi}'(\tau')
\nl&\quad\quad\quad
 \!+\!\int^t_{t_0}d\tau\int^{t}_{t_0}d\tau'
 \hat{\bm \xi}'^\dg(\tau)
\,{\bm {\wti g}}(\tau,\tau')\,
\hat{\bm \xi}(\tau')
\nl&\quad\quad\quad
\! +\!\int^t_{t_0}\!d\tau\int^{t}_{t_0}\!d\tau'
\hat{\bm \xi}'^\dg(\tau)
{\bm {\wti g}}(\tau,\tau')
\hat{\bm \xi}'(\tau')
 \!\Big\}.
\end{align}
Here we have introduced
$\hat{\bm \xi}^\dg(\tau)\equiv\left(\!
  \begin{array}{cc}
    \xi^\ast(\tau) ~~ \xi(\tau) \\
  \end{array}\!
\right)$ and $ \hat{\bm \xi}(\tau)\equiv\left(\!\begin{array}{c}
                                                        \xi(\tau)\\
                                                         \xi^\ast(\tau) \\
                                                       \end{array}\!
                                                     \right)$.
The matrix functions ${\bm   g} (\tau,\tau') $
and ${\bm{\wti g}} (\tau,\tau')$ are given by
\bsube\label{gmatrixs}
\begin{align}
      {\bm g}(\tau,\tau')&\!= \! \left(\!
                                  \begin{array}{cc}
                                    g_{\rm L}(\tau,\tau')\!+\!g_{\rm R}(\tau,\tau')
                                    & g_{\rm L}(\tau,\tau')\!-\!g_{\rm R}(\tau,\tau') \\
                                    g_{\rm L}(\tau,\tau')\!-\!g_{\rm R}(\tau,\tau')
                                    & g_{\rm L}(\tau,\tau')\!+\!g_{\rm R}(\tau,\tau') \\
                                  \end{array}\!
                                \right),
    \\
     {\bm  {\wti g} }(\tau,\tau')&\!= \! \left(\!
                                  \begin{array}{cc}
                                     g^+_{\rm L}(\tau,\tau')+g^+_{\rm R}(\tau,\tau')
                                    & g^+_{\rm L}(\tau,\tau')-g^+_{\rm R}(\tau,\tau') \\
                                    g^+_{\rm L}(\tau,\tau')-g^+_{\rm R}(\tau,\tau')
                                    & g^+_{\rm L}(\tau,\tau')+g^+_{\rm R}(\tau,\tau') \\
                                  \end{array}\!
                                \right),
\end{align}
\esube
where the correlation functions
$g_{\alpha}(\tau,\tau')$ and $g^+_{\alpha}(\tau,\tau')$ ($\alpha={\rm L, R}$)
have been given by \Eq{corr-bath} in the main text.

Noting that the path integral in \Eq{ppgf} is of the quadratic type,
it is thus possible to carry out the exact result.
Applying the variational calculus,
one can obtain the ``classical trajectory" (i.e. the extremal trajectory);
and the exponential factor of the path integral result
is given by the action determined by the ``classical trajectory". We have
\begin{align}\label{J1}
&\mathcal{J}(\xi^{\ast}_{f}, \xi'_{f};t|  \xi_{0}, \xi'^{\ast}_{0};t_0)
\nl
&={\cal N}(t) \exp\Big\{
\frac{1}{2}\big[ \xi^{\ast}_{f}\xi (t)+ \xi^{\ast}(t_0)\xi_{0}
+  \xi'^{\ast}(t) \xi'_{f}+  \xi'^{\ast}_{0} \xi'(t_0)\big]
\Big\},
\end{align}
where ${\cal N}(t)$ is the normalization factor
and $\xi(t),\xi^\ast(t_0),\xi'(t_0), \xi'^{\ast}(t)$ are determined by
the ``classical trajectory" which obeys the equations-of-motion
\bsube\label{sap0}
\begin{align}
\frac{d}{d\tau}\hat{\bm \xi}(\tau)
&+ i{\bm\varepsilon_{\M}}
\hat{\bm \xi}(\tau)
=
\!-\!\int_{t_0}^{\tau}d\tau'   {\cal G} (\tau,\tau')
\hat{\bm \xi}(\tau')
\nl&\quad
\!+\!\int_{t_0}^{t}d\tau' \wti{\cal G}(\tau,\tau')
           \big[\hat{\bm \xi}(\tau')+\hat{\bm \xi}'(\tau')\big],
\label{sapxi}
\\
\frac{d}{d\tau}\hat{\bm \xi}'(\tau)
&\!+\! i{\bm\varepsilon_{\M}}
\hat{\bm \xi}'(\tau)
=
 -\int_{t_0}^{\tau}d\tau' {\cal G}(\tau,\tau')
 \hat{\bm \xi}'(\tau')
\nl&
+\int_{t_0}^{t}d\tau' \wti{\cal G}(\tau',\tau)
\big[\hat{\bm \xi}(\tau')+\hat{\bm \xi}'(\tau')\big] \,.
\label{sapxiast}
\end{align}
\esube
In this result, we introduced
${\bm\varepsilon_{\M}}=\left(
  \begin{array}{cc}
    \varepsilon_{\M}  & 0 \\
    0 & -\varepsilon_{\M}  \\
  \end{array}
\right)$
and the $2\times2$ matrix functions
\bsube\label{calG-corr}
\begin{align}
 {\cal G}(\tau,\tau')&={\bm g}(\tau,\tau')+{\bm g}(\tau',\tau)
=2{\rm Re}[{\bm g}(\tau,\tau')],
\\
 \wti{\cal G}(\tau,\tau')&=
{\bm {\wti g}}(\tau,\tau')-{\bm {\wti g}}(\tau',\tau)+{\bm g}(\tau',\tau).
\end{align}
\esube
The solution of \Eq{sap0} can be formally expressed as \cite{Tu08235311,Lai18054508}
\bsube\label{xitrans}
\begin{align}
&\hat{\bm \xi}(\tau)
 ={\bm u}(\tau,t_0) \left(
                \begin{array}{c}
                  \xi_0  \\
                  \xi^{\ast}(t_0)\\
                \end{array}
              \right)
 +{\bm v}(\tau,t) \left(
                \begin{array}{c}
                  \xi(t)+\xi'_f  \\
                  \xi^{\ast}_f+\xi'^{\ast}(t)\\
                \end{array}
              \right),
 \label{xitaueq}
\\
&\hat{\bm \xi}(\tau)+\hat{\bm \xi}'(\tau)
 ={\bm u}^\dg(t,\tau)   \left(
                \begin{array}{c}
                  \xi(t)+\xi'_f  \\
                  \xi^{\ast}_f+\xi'^{\ast}(t)\\
                \end{array}
              \right),
\end{align}
\esube
while the two propagating factions ${\bm u}$ and ${\bm v}$
satisfy the following equations
\bsube\label{uveq}
\begin{align}
\label{ueq}
&\frac{d}{dt}{\bm u}(t,t_0)+i{\bm\varepsilon_{\M}}{\bm u}(t,t_0)
\!+\!\!\int_{t_0}^{t}\!d\tau {\cal G}(t,\tau) {\bm u}(\tau,t_0)\!=\!0,
\\
&\frac{d}{d\tau}{\bm v}(\tau,t)+i{\bm\varepsilon_{\M}}{\bm v}(\tau,t)
\!+\!\int_{t_0}^{\tau}\!d\tau'
{\cal G}(\tau,\tau') {\bm v}(\tau',t)
\nl&\quad\quad\quad
\!=\!\int_{t_0}^{t}\!d\tau\wti{\cal G}(\tau,\tau') {\bm u}^\dg(t,\tau'),
\label{veq}
\end{align}
\esube
under the end-points condition of
$\bm u(t,t_0)={\bf 1}$ and $\bm v(t_0,t)=0$.

With the help of \Eq{xitrans},
the $\mathcal{J}$-function can be reexpressed as
\begin{align}\label{ppgf2}
&\mathcal{J}(\xi^{\ast}_{f}, \xi'_{f};t|  \xi_{0}, \xi'^{\ast}_{0};t_0)
\nl&={\cal N}(t)\!\exp\!\Big\{\!
 \left(\!
   \begin{array}{cc}
     \xi^{\ast}_{f} & \xi'_f \\
   \end{array}\!
 \right)
\! {\bm J}_1(t)\!
 \left(\!
   \begin{array}{c}
     \xi_0 \\
      \xi'^{\ast}_0 \\
   \end{array}\!
 \right)
\!+\!
\left(\!
   \begin{array}{cc}
      \xi^{\ast}_{f} & \xi'_f \\
   \end{array}\!
 \right)
\! {\bm J}_2(t)\!
 \left(\!
   \begin{array}{c}
     \xi'_f \\
     \xi^{\ast}_f \\
   \end{array}\!
 \right)
 \nl&
 \!+\!
  \left(\!
   \begin{array}{cc}
     \xi'^{\ast}_{0} & \xi_0 \\
   \end{array}\!
 \right)
 \!{\bm J}_3(t)\!
 \left(\!
   \begin{array}{c}
     \xi_0 \\
      \xi'^{\ast}_0 \\
   \end{array}\!
 \right)
 \!+\!
  \left(\!
   \begin{array}{cc}
      \xi'^{\ast}_{0} & \xi_0 \\
   \end{array}\!
 \right)
 \!{\bm J}^\dg_1(t)\!
 \left(\!
   \begin{array}{c}
     \xi'_f \\
   \xi^{\ast}_f \\
   \end{array}\!
 \right)\!\Big\} \,.
\end{align}
Here, three matrix functions, ${\bm J}_i(t) (i=1,2,3)$, are introduced as
\bsube\label{Jmatrix}
\begin{align}
{\bm J}_1(t)&=\frac{1}{2a(t)}\left(
             \begin{array}{cc}
                2u_{11}(t) & -u_{12}(t)\\
               u_{12}(t) & 0 \\
             \end{array}
           \right),
\\
{\bm J}_2(t)&=\frac{1}{a(t)}\left(
             \begin{array}{cc}
                 1-a(t)  & 0\\
               0 & 0 \\
             \end{array}
           \right),
\\
{\bm J}_3(t)&=\frac{-1}{a(t)}\left(
             \begin{array}{cc}
                b(t) & 0\\
               0 & 0 \\
             \end{array}
           \right),
\end{align}
\esube
where
\bsube
\begin{align}\label{achit}
a(t)&= 1-v_{11}(t)-u_{12}(t)u^\ast_{21}(t),
\\
b(t)&=1-v_{11}(t)-u^\ast_{11}(t)u_{11}(t) .
\end{align}
\esube
In the above results, for simplicity, we have denoted
$u_{ij}(t)\equiv u_{ij}(t,t_0)$ and $v_{ij}(t)\equiv v_{ij}(t,t )$.

\subsection{Path-integral formulation for the propagation\\
of the ADOs $\rho^{\pm}_{\alpha }(t)$}

Following the above treatment for the evolution of $\rho(t)$,
we can formulate the evolution of $\rho^{\pm}_{\alpha }(t)$
in terms of path integral as follows
\begin{align}\label{rdmaux}
\langle\xi_f|\rho^{\pm}_{\alpha}(t)|\xi'_f\rangle&=
\int d\mu(\xi_0) d\mu(\xi'_0)
\langle\xi_0|\rho(t_0)|\xi'_0\rangle \nonumber
\\ &
\times {\cal J}^{\pm}_{\alpha}(\xi^\ast_f,\xi'_f,t|\xi_0,\xi'^{\ast}_0,t_0),
\end{align}
where the propagating function is given by the path integral as
\begin{align}\label{Appgf}
{\cal J}^{\pm}_{\alpha}(\xi^{\ast}_f,\xi'_f,t|\xi_0, \xi'^{\ast}_0,t_0)
\!=\!\!
\int\!\mathcal{D}[ \bm\xi; \bm\xi']
e^{iS_{\rm eff}[ \bm\xi;\bm\xi']}
\mathcal{F}^{\pm}_{\alpha}
[\bm\xi;\bm\xi'].
\end{align}
Based on the result of $\mathcal{F}[\bm\xi;\bm\xi']$,
applying the technique of generating functional, we can obtain
$  {\cal F}^\pm_{\alpha }[ \bm\xi;\bm\xi']
 = {\cal F}^\pm_{\alpha 1}[ \bm\xi;\bm\xi']
 +{\cal F}^\pm_{\alpha 2}[\bm\xi;\bm\xi']$,
with the two parts given by
\bsube
\begin{align} \label{AIFs}
  {\cal F}^\pm_{\alpha 1 }[\bm\xi;\bm\xi']
&=-i{\cal Q}^{\pm}_{\alpha 1 }(t;\{\xi,\xi',\xi^\ast,\xi'^\ast\}) {\cal F}[ \bm\xi;\bm\xi'],
\\
{\cal F}^{\pm}_{\alpha 2}[ \bm\xi;\bm\xi']
&=-i {\cal Q}^{\pm}_{\alpha 2}(t;\{\xi,\xi',\xi^\ast,\xi'^\ast\}) {\cal F}[ \bm\xi;\bm\xi'].
\end{align}
\esube
More explicitly, the pre-factors ${\cal Q}^{\pm}_{\alpha 1,2}$ read as
\bsube
\begin{align*}
{\cal Q}^{+}_{\alpha 1}(t;\{ \xi^\ast, \xi'^\ast\})
&=
- \int_{t_0}^{t}\!d\tau\,
     g^{\ast}_{\alpha }(t-\tau)     \xi'^\ast(\tau)
\nl&\quad
  +  \int_{t_0}^{t}\!d\tau\,
   g^{+\ast}_{\alpha }(t-\tau)
     \left[\xi^\ast(\tau)+\xi'^\ast(\tau)\right],
\\
 {\cal Q}^{+}_{\alpha 2}(t;\{\xi,\xi'\})
&=
-(-)^{\alpha} \int_{t_0}^{t}\!d\tau\,
     g^{\ast}_{\alpha }(t-\tau)     \xi'(\tau)
\nl&
 + (-)^{\alpha}  \int_{t_0}^{t}\!d\tau\,
   g^{+\ast}_{\alpha }(t-\tau)
     \left[\xi(\tau)+\xi'(\tau)\right],
\\
{\cal Q}^{-}_{\alpha 1}(t;\{\xi,\xi'\})&=
 \int_{t_0}^{t}\!d\tau\,
     g_{\alpha }(t-\tau)   \xi(\tau)
 \nl&
 -  \int_{t_0}^{t}\!d\tau\,
   g^+_{\alpha}(t-\tau)
    \left[\xi(\tau) +\xi'(\tau)\right],
\\
 {\cal Q}^{-}_{\alpha 2}(t;\{\xi^\ast,\xi'^\ast\})&=(-)^{\alpha}
 \int_{t_0}^{t}\!d\tau\,
     g_{\alpha }(t-\tau)   \xi^\ast(\tau)
 \nl&
  - (-)^{\alpha} \int_{t_0}^{t}\!d\tau\,
   g^+_{\alpha}(t-\tau)
    \left[\xi^\ast(\tau) +\xi'^\ast(\tau)\right].
\end{align*}
\esube
Since the path integral in \Eq{Appgf} is also of the quadratic form,
we can substitute the ``classical trajectory" of \Eq{xitrans} into the action
of the path integral, which yields the same exponential factor of
${\cal J}_{\alpha}(\xi^{\ast}_f,\xi'_f,t|\xi_0, \xi'^{\ast}_0,t_0)$,
while the ${\cal Q}^{\pm}_{\alpha 1,2}$-factors
can be expressed in terms of the ``end-point-coordinates" as
 \begin{widetext}
\bsube
\begin{align}
{\cal Q}^{-}_{\alpha 1}(t;\{\xi_0,\xi'^\ast_0,\xi'_f,\xi^\ast_f\})&=
\wti X_{\rm \alpha e11}(t) \xi_0
+\wti X_{\alpha e12}(t)\xi'^\ast_0
+\wti Y_{\alpha e11}(t) \xi'_f
+\wti Y_{\alpha e12}(t)\xi^\ast_f
\\
{\cal Q}^{-}_{\alpha  2}(t;\{\xi_0,\xi'^\ast_0,\xi'_f,\xi^\ast_f\})&=
-(-)^\alpha\left[\wti X_{\alpha h11}(t) \xi'^\ast_0
+\wti X_{\alpha h12}(t)\xi_0
+\wti Y_{\alpha h11}(t) \xi^\ast_f
+\wti Y_{\alpha h12}(t)\xi'_f
\right]
\\
{\cal Q}^{+}_{\alpha  1}(t;\{\xi_0,\xi'^\ast_0,\xi'_f,\xi^\ast_f\})
&=
-\left[\wti X^\ast_{\alpha e11}(t)\xi'^\ast_0
+\wti X^\ast_{\alpha e12}(t) \xi_0
+\wti Y^\ast_{\alpha e11}(t) \xi^\ast_f
+\wti Y^\ast_{\alpha e12}(t) \xi'_f
\right]
\\
 {\cal Q}^{+}_{\alpha  2}(t;\{\xi_0,\xi'^\ast_0,\xi'_f,\xi^\ast_f\})
&=(-)^\alpha\left[\wti X^\ast_{\alpha h11}(t) \xi_0
+\wti X^\ast_{\alpha h12}(t)\xi'^\ast_0
+\wti Y^\ast_{\alpha h11}(t) \xi'_f
+\wti Y^\ast_{\alpha h12}(t)\xi^\ast_f
\right]
\end{align}
\esube
\end{widetext}
In this result, the coefficients
$\widetilde{X}_{\alpha e(h)ij}$ and $\widetilde{Y}_{\alpha e(h)ij}$
have lengthy expressions in terms of the $u_{ij}$ and $v_{ij}$ functions,
being thus not shown here.

\subsection{Operator forms of $\rho^{\pm}_{\alpha }(t)$}

In order to extract out the operator forms of $\rho^{\pm}_{\alpha }(t)$
from the path-integral results of
${\cal J}(\xi^{\ast}_f,\xi'_f,t|\xi_0, \xi'^{\ast}_0,t_0)$ and
${\cal J}^{\pm}_{\alpha}(\xi^{\ast}_f,\xi'_f,t|\xi_0, \xi'^{\ast}_0,t_0)$,
we need to express the results of ${\cal Q}^{\pm}_{\alpha 1,2}$
using only the {\it final end-point-coordinates}, i.e.,
${\cal Q}^{\pm}_{\alpha 1,2}[\xi^{\ast}_f; \xi'_f]$.
This can be achieved as follows.
Making derivatives on ${\cal J}(\xi^{\ast}_f,\xi'_f,t|\xi_0, \xi'^{\ast}_0,t_0)$,
i.e., $\frac{\partial {\cal J}}{\partial \xi^\ast_f}$
and $\frac{\partial {\cal J}}{\partial\xi'_f}$,
from their results we then carry out the following expressions:
\bsube\label{xi0calJ}
\begin{align}
\xi_0{\cal J}&=\frac{a}{c}u^\dg_{11}
\frac{\partial {\cal J}}{\partial \xi^\ast_f}
-\frac{a}{c}u_{12}
\frac{\partial {\cal J}}{\partial\xi'_f}  \nl
&
-\frac{1-a}{c} u^\dg_{11}\xi'_f{\cal J}
-\frac{1-a}{c}u_{12} \xi^\ast_f{\cal J}  \,,
 \\
  \xi'^\ast_0{\cal J}&=\frac{a}{c}u^\dg_{12}
\frac{\partial {\cal J}}{\partial \xi^\ast_f}
-\frac{a}{c}u_{11}
\frac{\partial {\cal J}}{\partial\xi'_f} \nl
&
-\frac{1-a}{c}u^\dg_{12}\xi'_f{\cal J}
-\frac{1-a}{c}u_{11} \xi^\ast_f{\cal J}  \,.
\end{align}
\esube
Here, in addition to $a(t)$ and $b(t)$ as shown in \Eq{achit},
we further introduced $c(t)=u_{11}(t)u^\dg_{11}(t)-u_{12}(t)u^\ast_{12}(t)$.

Then, let us reexpress the path integral result of $\rho^{\pm}_{\alpha}(t)$ as
\begin{align}\label{ADO-ff}
& \langle\xi_f|\rho^{\pm}_{\alpha}(t)|\xi'_f\rangle
 = \int d\mu(\xi_0) d\mu(\xi'_0)
\langle\xi_0|\rho(t_0)|\xi'_0\rangle \nonumber
\\ &
~~ \times \left(\sum_{j=1,2} (-i) {\cal Q}^{\pm}_{\alpha j}[\xi^{\ast}_f; \xi'_f]
{\cal J}(\xi^\ast_f,\xi'_f,t|\xi_0,\xi'^{\ast}_0,t_0) \right)    \nl
& ~~ = \sum_{j=1,2} (-i) {\cal Q}^{\pm}_{\alpha j}[\xi^{\ast}_f; \xi'_f]
\langle\xi_f|\rho(t)|\xi'_f\rangle   \,.
\end{align}
Based on this result, using the following properties
of the fermionic coherent state
\begin{align}\label{dalgebra}
  \xi|\xi\ra&=f|\xi\ra,~~~  -\frac{\partial}{\partial\xi} |\xi\ra=f^\dg|\xi\ra,\\
  \la\xi|\xi^\ast&= \la\xi|f^\dg,~~~  \la\xi|\frac{\partial}{\partial\xi^\ast}=\la\xi|f \,,
\end{align}
we can extract out the operator forms of $\rho^{\pm}_{\alpha}(t)$.
For instance, let us consider the term in \Eq{ADO-ff}
\begin{align}
\int d\mu(\xi_0) d\mu(\xi'_0)
& ~ \langle\xi_0|\rho(t_0)|\xi'_0\rangle
\frac{\partial}{\partial \xi^\ast_f}
{\cal J}(\xi^\ast_f,\xi'_f,t|\xi_0,\xi'^{\ast}_0,t_0)  \nl
&= \frac{\partial}{\partial \xi^\ast_f} \langle\xi_f|\rho(t)|\xi'_f\rangle  \nl
&= \langle\xi_f|f \rho(t)|\xi'_f\rangle \,,
\end{align}
which allows us to extract the term of ``$f \rho(t)$" for $\rho^{\pm}_{\alpha}(t)$.
Other terms in \Eq{ADO-ff} can be handled similarly.
Splitting the results of $\rho^{\pm}_{\alpha}(t)$ into two parts,
i.e., $\rho^{\pm}_{\alpha}(t)=\sum_{j=1,2}\rho^{\pm}_{\alpha j}(t)$,
we finally obtain
\bsube\label{ADOsoper}
\begin{align}
\rho^{-}_{\alpha 1}(t )
&=
-i\big\{ \big[
\kappa_{\alpha11}(t)-\lambda _{\alpha11}(t)\big]
  f\rho(t)
  \nl&\quad
  - \big[
 \kappa_{\alpha12}(t)- \lambda _{\alpha12}(t) \big]
\rho(t)  f^\dg
  \nl&\quad
  -\lambda _{\alpha11}(t)   \rho(t)   f
 + \lambda _{\alpha 12}(t)    f^\dg\rho(t)\big\},
  \\
\rho^{-}_{\alpha  2}(t)
&=-(-)^{\alpha}
i\big\{ \big[
 \kappa_{\alpha 21}(t)- \lambda _{\alpha 21}(t)\big]
 f\rho(t)
 \nl&\quad- \big[
\kappa_{\alpha 22}(t)-\lambda _{\alpha 22}(t) \big]
\rho(t)  f^\dg
 \nl&\quad- \lambda _{\alpha 21}(t)   \rho(t)   f
 +\lambda _{\alpha 22}(t)    f^\dg\rho(t)\big\} \,.
\end{align}
\esube
For $\rho^{+}_{\alpha j}(t)$, one can easily prove that
$\rho^{+}_{\alpha j}(t)=[\rho^{-}_{\alpha j}(t)]^\dg$.
The time-dependent coefficients in the above results
are the matrix elements of
${\bm\kappa_{\alpha}}(t)$ and ${\bm\lambda_{\alpha}}(t)$,
while the latter are given by
\bsube\label{kap-lamb}
\begin{align}
{\bm\kappa}_{\alpha }(t) &\!=\! \int_{t_0}^{t}\!d\tau\,g_{\alpha }(t-\tau)
{\bm u}(\tau) [{\bm u}(t)]^{-1},
\end{align}
\begin{align}
{\bm\lambda}_{\alpha }(t) &\!= \!\!\int_{t_0}^{t}\!d\tau\,g_{\alpha }(t-\tau)
\left(\!
  \begin{array}{cc}
    a(t) u_{11}(\tau) & b(t) u_{12}(\tau) \\
     b(t) u^\ast_{12}(\tau) & a(t) u^\ast_{11}(\tau) \\
  \end{array}\!
\right)\!
  [{\bm u}(t)]^{-1}
  \nl&\!-\!
  \int_{t_0}^{t}\!d\tau\,g_{\alpha }(t-\tau)
\left(\!
  \begin{array}{cc}
      v_{11}(\tau) &  -v_{12}(\tau) \\
     -v^\ast_{12}(\tau) &   v^\ast_{11}(\tau) \\
  \end{array}\!
\right)
\nl&
\!+\!\!\int_{t_0}^{t}\!d\tau
\left(\!\!
  \begin{array}{cc}
      g^+_{\alpha }(t-\tau) \!&\!  0 \\
     0 \!&\!  {g}^-_{\alpha }(t-\tau) \\
  \end{array}\!\!
\right)\!
\left(\!\!
  \begin{array}{cc}
      \bar{u}^\ast_{11}(\tau) \!&\!  -\bar{u}^\ast_{12}(\tau) \\
     -\bar{u}_{12}(\tau) \!&\!   \bar{u}_{11}(\tau) \\
  \end{array}\!\!
\right).
\end{align}
\esube
Under the assumption of wide-band limit,
which makes $g_{\alpha}(t-\tau)\rightarrow\Gamma_{\alpha}\delta(t-\tau)$,
we can further simplify the results of \Eq{kap-lamb} as follows:
\bsube
\begin{align*}
\kappa_{\alpha 11}(t)&=\kappa_{\alpha 22}(t)=\int_{t_0}^{t}\!d\tau\,g_{\alpha }(t-\tau),
\\
 \kappa_{\alpha 12}(t)&=\kappa_{\alpha 21}(t)=0,
\\
 \lambda_{\alpha 11}(t)
 &=
 \int_{t_0}^{t}\!d\tau\, g^+_{\alpha }(t-\tau)u^\dg_{11}(t,\tau),
\\
 \lambda_{\alpha 12}(t)
 &=
 \int_{t_0}^{t}\!d\tau\,g^+_{\alpha }(t-\tau)u^\dg_{12}(t,\tau),
\\
 \lambda_{\alpha 21}(t)
 &=
 \int_{t_0}^{t}\!d\tau\, {g}^-_{\alpha }(t-\tau)u_{12}(t,\tau),
\\
\lambda_{\alpha 22}(t)
  &=
 \int_{t_0}^{t}\!d\tau\,  {g}^-_{\alpha }(t-\tau)u_{11}(t,\tau) \,.
\end{align*}
\esube
These are the results we used in the main text.

So far, we have established the relationship between
$\rho^{\pm}_{\alpha}(t)$ and $\rho(t)$, as shown in \Eq{ADOsoper},
together with the various coefficients as shown above.
Then, straightforwardly, substituting the result of
$\rho^{\pm}_{\alpha}(t)$ given by \Eq{ADOsoper} into \Eq{LvEq},
we obtain \Eq{EME-wbl} in the main text,
i.e., the central result of Majorana master equation analyzed in this work.
Beyond the wide-band limit, the various rates in the master equation read as
\bsube\label{coeffsm1-gen}
\begin{align}
{\bm\Gamma^{+}_{\alpha}}(t)
&\!=\!2{\rm Re}\!
\left(\!
  \begin{array}{cc}
    \lambda_{\alpha11} (t)& \lambda_{\alpha12} (t)\\
    \kappa_{\alpha21}(t)\!-\!\lambda_{\alpha21}(t) & \kappa_{\alpha22}(t)\!-\!\lambda_{\alpha22}(t) \\
  \end{array}\!\!
\right),
\\
{\bm\Gamma^{-}_{\alpha}}(t)
&\!=\!2{\rm Re}\!
\left(\!\!
  \begin{array}{cc}
    \kappa_{\alpha11}(t)\!-\!\lambda_{\alpha11} (t)& \kappa_{\alpha12}(t)\!-\!\lambda_{\alpha12} (t)\\
    \lambda_{\alpha21}(t) &  \lambda_{\alpha22}(t) \\
  \end{array}\!\!
\right),   \\
\Upsilon(t) &=\sum_{\alpha}\Big\{(-)^\alpha
\big[\kappa_{\alpha 11}(t)+\kappa^\ast_{\alpha 22}(t)\big]
\nl&\quad
+ \big[ \kappa^\ast_{\alpha 12}(t)+ \kappa_{\alpha 21}(t)\big]\Big\}.
\end{align}
\esube
Under the assumption of wide-band limit,
these results are simplified to \Eq{rates-wbl} in the main text.

For the transport currents, let us reexpress \Eq{curr1} as
 \bsube\label{currt2}
\begin{align}
I_{\alpha 11}(t)&= i\frac{e}{\hbar}
 {\rm tr}_{\s} \big[ \rho^+_{\alpha1}(t)f -f^\dg\rho^-_{\alpha1}(t) \big],
 \\
 I_{\alpha 22}(t)&= i\frac{e}{\hbar}
 {\rm tr}_{\s}
 \big[\rho^+_{\alpha 2}(t)  f^\dg-   f\rho^-_{\alpha 2}(t)\big] ,
 \\
   I_{\alpha 12}(t)&= i\frac{e}{\hbar}
 {\rm tr}_{\s} \big[\rho^+_{\alpha 1}(t)  f^\dg-  f\rho^-_{\alpha 1}(t)\big],
 \\
 I_{\alpha 21}(t)&=  i\frac{e}{\hbar}
 {\rm tr}_{\s}
 \big[\rho^+_{\alpha 2}(t)  f-   f^\dg\rho^-_{\alpha 2}(t)\big].
\end{align}
\esube
Again, straightforwardly, substituting the result of $\rho^{\pm}_{\alpha}(t)$
into the above expressions, we obtain the current formulas \Eq{curr-rate} in the main text.

\end{document}